# Modulation Classification of MIMO-OFDM Signals by Independent Component Analysis and Support Vector Machines

Yu Liu, A.M. Haimovich, *Fellow, IEEE*, Wei Su, Jason Dabin, Emmanuel Kanterakis

*Abstract*—A modulation classification (MC) scheme based on Independent Component Analysis (ICA) in conjunction with either maximum likelihood (ML) or Support Vector Machines (SVM) is proposed for MIMO-OFDM signals over frequency selective, time varying channels. The method is blind in the sense that it is assumed that the receiver has no information about the channel and transmitted signals other than that the spatial streams of signals are statistically independent. The processing consists of separation of the MIMO streams followed by modulation classification of the separated signals. While in general, blind separation of signals over frequency selective channels is a difficult problem, the non-frequency selective nature of the channel experienced by individual symbols in a MIMO-OFDM system enables the application of well-known ICA algorithms. Modulation classification is implemented by maximum likelihood and by an SVM-based modulation classification method relying on pre-selected modulation-dependent features. To improve performance in time varying channels, the invariance of the channel is exploited across the coherence bandwidth and the time coherence. The proposed method is shown to perform with high probability of correct classification over realistic ITU pedestrian and vehicular channels. An upper bound on the probability of correct classification is developed based on the Cramer Rao bound of channel estimation.

*Index Terms*—Modulation classification, Cramer-Rao lower bounds, maximum likelihood, MIMO OFDM.

## I. INTRODUCTION

TO meet the growing demand for high-data rates in communications systems, new wireless applications rely on multiple-input multiple-output (MIMO) technologies. MIMO can support increased data capacity through, spatial multiplexing, i.e., the transmission of data in parallel streams. Orthogonal Frequency Division Multiplexing (OFDM) is a multicarrier transmission technique where the frequency band is divided into several orthogonal sub-bands, such that the symbols transmitted on each sub-band experience frequency non-selective fading. Channel equalization is then reduced to a one-tap filter per data symbol. The combination of MIMO transmission and OFDM data modulation is central to fourth generation (4G) wireless technologies, such as WiMax, LTE, and IEEE 802.22.

Recognition of the modulation of unknown received signals has obvious military applications. As for civilian applications, attempts to reduce overhead of reference signals required for channel estimation has motivated research in blind and semi-blind MIMO techniques. Blind techniques are also expected to play a role in software defined radio and cognitive radio. Configuration information required by a software defined radio system is transmitted as overhead to the data. However, intelligent receivers capable of extracting this information blindly may improve transmission efficiency through reductions in overhead. For example, automatic modulation classification eliminates the need for supplementary information on the modulation type.

In general, classification requires preprocessing of the received signals for acquiring signal parameters, such as carrier frequency and symbol rate. This paper focuses on the modulation classification of MIMO-OFDM signals assuming that frequency and time synchronization have already been attained.

Modulation classification methods for single-input single-output (SISO) systems are generally classified as likelihood-based [1], [15], [16], [27]-[36] or feature-based [1], [10], [11], [37]-[45]. A detailed survey of automatic modulation classification methods for SISO systems is given in [1]. Likelihood-based methods rely on the computation of the likelihood function of the received signal under different hypotheses on the modulation types. Depending on ways of coping with unknown quantities, which include unknown symbol constellation points and unknown channel, various modulation classification techniques have been proposed in the literature. Among likelihood-based methods, an approach referred to as hybrid likelihood ratio test (HLRT) [36], [32] has been shown to provide a promising solution to the modulation classification problem in SISO systems. It computes a metric based on blind channel estimation and on averaging the likelihood function over symbol constellation points. HLRT is capable of recognizing some common modulation types with reasonable computational complexity. Our proposed likelihood-based classification method extends the HLRT-based approach to MIMO OFDM.

Likelihood-based classification is optimal in the sense of attaining minimum probability of misclassification, but is computationally complex, often requiring exhaustive searches through unknown parameters. With feature-based methods, specific features are extracted from the signal and compared with pre-calculated values. Feature-based methods are usually ad-hoc, but computationally efficient. A variety of feature-based methods have been developed based on different selected features. In this paper, we adopt cumulant-based features [10] due to their ability to recognize a number of modulations [1], high correct classification rate and robustness with respect to inaccurate carrier phase estimates.

MIMO modulation classification is particularly challenging due to the interference between received signals and the multiplicity of unknown channels. Modulation classification of MIMO signals relies on the blind estimation of the MIMO channel. Blind MIMO channel estimation has been an active area of research (e.g. [2] and [3]). MIMO-OFDM blind channel estimation has been studied in [4]. Independent component analysis (ICA) [6] is a class of blind source separation (BSS) methods for separating linear mixtures of signals into independent components. ICA can recover signals from a mixture, up to certain ambiguities, if the signals are statistically independent and non-Gaussian. ICA can also be viewed as a solution to the blind channel estimation problem, when the MIMO channel is frequency-flat and time-invariant. Computationally efficient algorithms have been developed for ICA, which encouraged their application to large-scale problems. A likelihood-based approach to MIMO modulation classification is proposed in [5], where the channel matrix required for the calculation of the likelihood is first estimated blindly by ICA.

One of the goals of this paper is to develop a theoretical performance analysis of modulation classification in MIMO OFDM systems. In MIMO OFDM, communication takes place over parallel, flat fading channels. A theoretical bound on the performance of modulation classification in SISO channels has been developed in [15]. To the best of our knowledge, such a performance bound for MIMO systems has not yet been shown. Our approach is to develop an upper bound on the probability of correct classification (PCC) from the CRB of the estimates of the flat MIMO channel. With the blind MIMO problem, channel estimation is hampered by interference from other MIMO channels as well as noise. Various CRB's for the blind, but noiseless, real MIMO channel with continuous source variables are presented in [21]-[26]. Unlike the references, the CRB we propose in the current paper is for the discrete source variables, such as those found in digital communication systems. Additionally, our proposed CRB accounts for the effect of white Gaussian noise. Since popular modulation formats, such as QPSK and QAM, involve complex signals, the CRB calculation is performed for the complex case. Modulation classification based on the channel CRB is compared with the performance with perfect channel knowledge and with numerical results yielded by combining maximum likelihood (ML) classification and ICA channel estimation.

The main contributions of the current paper are: (1) exploit the frequency non-selective channel experienced by the MIMO-OFDM data symbols and the finite frequency and time selectivity to perform modulation classification on groups of data symbols with a common channel; (2) develop a low complexity SVM-based modulation classifier; (3) develop a CRB of flat MIMO channel estimates for both data-aided and blind channel estimation; (4) propose an upper bound on the performance of modulation classification over flat MIMO channels.

The rest of the paper is organized as follows: the next section introduces the signal model, the proposed MIMO-OFDM modulation classification methods are presented in Section III, the CRB for the estimation of MIMO channel with frequency-flat fading and an upper bound on modulation classification are developed in Section IV, numerical examples are provided in Section V, and Section VI wraps up with conclusions.

Notations: Notations: the notation $(\cdot)^T$ denotes transpose; $(\cdot)^H$ denotes Hermitian operation; $(\cdot)^+$ denotes the pseudoinverse of a matrix; the superscripts ($R$) and ($I$) denote respectively, the real part and the imaginary part of a complex number or a complex matrix.

## II. SIGNAL MODEL

Consider a MIMO-OFDM system with $M_t$ transmit antennas and $M_r$ receive antennas. Identifiability conditions of the MIMO channel require, $M_t \leq M_r$. The system transmits frames of OFDM symbols $\mathbf{s}^{(i)}(k,n)$, where $\mathbf{s}^{(i)}$ is a length $M_t$ vector of symbols belonging to a constellation $\Omega_i$, $n$ is the subcarrier index $n$ and $k$ is the frame index. A frame is an OFDM block of data symbols. The transmitted symbols are of unknown PSK/QAM modulation, but are assumed statistically independent between antennas, subcarriers and frames. In addition, ideal time synchronization as well as ideal carrier frequency synchronization is assumed at the receiver side. A block diagram of the MIMO-OFDM system is shown in Fig. 1.

Assuming a cyclic prefix that ensures inter-carrier interference-free observations, the received length $M_r$ vector in the frequency domain, $\mathbf{y}(k,n)$, is expressed

$$\mathbf{y}(k,n) = \mathbf{H}(k,n)\mathbf{s}(k,n) + \mathbf{z}(k,n), \quad (1)$$

where $\mathbf{H}(k,n)$ is the MIMO channel matrix associated with subcarrier index $n$ and frame index $k$, and $\mathbf{z}(k,n)$ is additive white Gaussian noise. The noise is complex-valued, zero mean, has known variance $\sigma^2/2$ for both real and imaginary parts, and is independent between receive antennas, subcarriers, and frames.

A MIMO-OFDM system can be considered a set of instantaneous mixtures of transmitted signals. The problem of separating MIMO-OFDM signals becomes a blind source separation problem (BSS) at each subcarrier. But rather than having to solve multiple BSS problems, we exploit the coherence bandwidth and time coherence of the channel, assumed known at the receiver, to form a set of $K$ frames and $N$ subcarriers over which the channel is fixed and the same, i.e., $\mathbf{H}(k,n) = \mathbf{H}$ for $n=1,...,N$ and $k=1,...,K$. Note that assuming that the number of subcarriers of the OFDM frame matches the coherence bandwidth of the channel, implies a flat channel. The model can be easily expanded to a frequency selective channel by repeated estimation of the channel at the different coherence bandwidths. For notational convenience, the $KN$ observation vectors $\mathbf{y}(k,n)$, signal vectors $\mathbf{s}(k,n)$, and noise vectors $\mathbf{z}(k,n)$ associated with channel $\mathbf{H}$, are re-indexed $\mathbf{y}_k$, $\mathbf{s}_k$, and $\mathbf{z}_k$, respectively, for $k=1,\ldots,KN$. With that, the received signal model is

$$\mathbf{y}_k = \mathbf{H}\mathbf{s}_k + \mathbf{z}_k. \quad (2)$$

For future use, denote $\mathbf{Y}=\{\mathbf{y}_k\}_{k=1}^{KN}$ the set of all observations and $\mathbf{S}^{(i)}=\{\mathbf{s}_k\}_{k=1}^{KN}$ the set of all transmitted vectors belonging to constellation $\Omega_i$. Furthermore, without loss of generality, $\Omega_i$ is assumed to be a constellation of symbols with average power equal to 1.

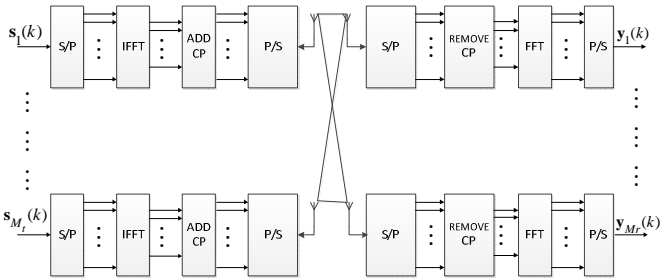

Fig. 1. MIMO-OFDM system model

### III. PROPOSED METHODS

The proposed method for blind classification of MIMO-OFDM signals has three stages as summarized in Fig. 2. The first stage groups subcarriers and frames to maximize the number of observations for a fixed channel matrix. The second stage applies an ICA algorithm to estimate the inverse channel matrix $\mathbf{H}^{-1}$ and thus separate the MIMO signals. Finally, in the third stage, modulation classification methods are applied to the separated data streams. The ICA and modulation classification stages are discussed below in detail.

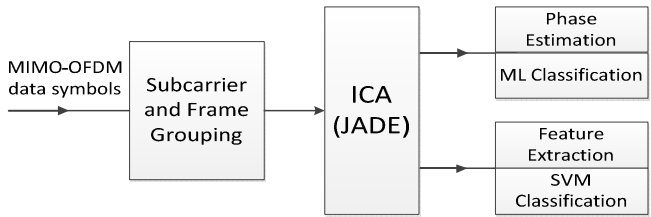

Fig. 2. Block diagram of MIMO-OFDM modulation classification

#### A. MIMO Separation by ICA

Modulation classification of MIMO signals relies on the blind estimation of the MIMO channel and blind source separation of signals. Reconstructing the unobserved channel matrix $\mathbf{H}$ and source signals $\mathbf{s}_k$ up to certain ambiguities from their linear mixtures $\mathbf{y}_k$, is possible if the source signals are statistically independent and non-Gaussian. ICA is a class of techniques for achieving that. Given the signal model (2), ICA methods [6-8] seek to compute a demixing matrix $\mathbf{W}=\mathbf{H}^{-1}$ by utilizing the independency between source signals. Since $\mathbf{W}=\mathbf{H}^{-1}$, computation of the demixing matrix is equivalent to the estimation of the channel matrix. Application of $\mathbf{W}$ to the data vector $\mathbf{y}_k$ enables to recover the vector of transmitted symbols $\mathbf{s}_k$, according to $\hat{\mathbf{s}}_k = \mathbf{W}\mathbf{y}_k$.

ICA estimates the channel matrix up to a permutation and phase ambiguity. In other words, ICA estimates the channel matrix

$$\hat{\mathbf{H}} = \mathbf{H}\mathbf{P}\mathbf{D} \quad (3)$$

where $\hat{\mathbf{H}}$ is the channel estimate obtained by ICA, $\mathbf{D}$ is a diagonal, complex-valued matrix accounting for phase ambiguities, and $\mathbf{P}$ is a permutation matrix. Due to phase and permutation ambiguities, a channel estimate obtained by using ICA cannot be used directly in the modulation classification algorithms.

A number of algorithms are available for implementing ICA. They have in common the aim to minimize a contrast function, which measures the statistical dependency of the sources in the mixture. Minimization of the contrast function yields the desired independent source signals and channel matrix. In this paper, we use the JADE algorithm [7, 8] due to its relative fast speed of convergence.

In practice, due to data support and noise limitations, the estimated channel matrix $\hat{\mathbf{H}}$ is only an approximation, apart from the ambiguties, of the true channel matrix. The accuracy of estimated channel matrix affects the modulation classification. Based on the accuracy of the estimated channel, an upper bound on the performance of modulation classification will be obtained later in this paper.

#### B. Maximum Likelihood Modulation Classification

For the MIMO-OFDM system (2), under the assumption of statistically independent received samples, the likelihood function of the observations is given by

$$L^{(i)}\left(\mathbf{Y}\mid\mathbf{S}^{(i)},\mathbf{H}\right)$$
$$= \frac{1}{\left(\pi\sigma^2\right)^{NK\cdot M_r}}\prod_{k=1}^{NK}\exp\left[-\frac{1}{\sigma^2}\left\|\mathbf{y}_k-\mathbf{H}\mathbf{s}_k\right\|^2\right] \quad (4)$$

where the norm is Euclidean. The likelihood is conditioned on the transmitted symbols $\mathbf{S}^{(i)}$ and channel $\mathbf{H}$, which are unknown. These unknown quantities are addressed in different ways. The ICA processing yields an estimate $\hat{\mathbf{H}}$ of the channel matrix, which is then used in (4). The difficulty posed by the unknown symbols $\mathbf{S}^{(i)}$ is resolved by assuming a uniform a priori distribution, and averaging over the symbols from each constellation $\Omega_i$. With these modifications, the likelihood function for constellation $\Omega_i$ is given by

$$L^{(i)}\left(\mathbf{Y}\mid\hat{\mathbf{H}}\right) = E_{\Omega_i}\left[L^{(i)}\left(\mathbf{Y}\mid\mathbf{S}^{(i)},\hat{\mathbf{H}}\right)\right] \quad (5)$$

The ICA stage of the processing produces a demixing matrix $\mathbf{W}$ or equivalently, an estimate of the channel matrix $\hat{\mathbf{H}} = \mathbf{W}^{-1}$. As discussed previously and indicated by (3), the channel estimate produced by ICA has inherent permutation and phase indeterminacies. It is shown in [5] that the likelihood function (5) is invariant to a permutation matrix $\mathbf{P}$ applied to $\mathbf{H}$, i.e. $L^{(i)}(\mathbf{Y}/\mathbf{H}) = L^{(i)}(\mathbf{Y}/\mathbf{HP})$.

Unlike its invariance to signal permutations, the likelihood function (5) is dependent on the unknown phase offsets contained in the diagonal of the matrix $\mathbf{D}$. These phase offsets need to be estimated for correct modulation classification. If the unknown phase offset for one of the separated MIMO streams is $\phi_m$, $m = 1,...,M_t$, the log likelihood function for estimating it is

$$\ell(\phi_m) = \sum_{k=1}^{KN} \ln\left\{\sum_{\Omega^{(i)}} \exp\left[-\frac{1}{2\sigma_x^2}\left|x_m(k) - s_m(k)e^{j\phi_m}\right|^2\right]\right\} \quad (6)$$

where $x_m(k)$ is a data symbol at the output of the ICA algorithm, $s_m(k)$ is one of the $M_t$ transmitted symbols at time $k$, and $\sigma_x^2$ is the interference term (multiuser interference due to imperfect MIMO channel equalization by the ICA and Gaussian noise). In [9] it is shown that as the SNR tends to zero, the ML phase estimator takes on the blind form

$$\hat{\phi}_m = \frac{1}{P} \arg\left[E\left[(s_m^*)^P\right] \sum_{k=1}^{KN} (x_m(k))^P\right] \quad (7)$$

where $P$ is the number of symbols of the constellation that are rotationally symmetric. For example, for QAM constellations, $P = 4$. According to [9], even after the phase offset estimation according to (7), there remains a leftover phase ambiguity corresponding to a multiple of the phase difference between the constellation symbols. It is an easy argument to make that this remaining ambiguity does not interfere with the modulation classification. For future use, let $\mathbf{D}_1$ denote the diagonal matrix formed by the elements $e^{j\phi_m}$.

Substituting the estimated channel matrix $\hat{\mathbf{H}}$ multiplied by the inverse of the phase offsets matrix $\mathbf{D}_1$ in the likelihood expression, and averaging over the symbols of each hypothesized constellation, leads to the following likelihood-based modulation classification

$$\hat{\Omega} = \arg\max_{\Omega_i} \ln\left\{L^{(i)}\left(\mathbf{Y}/\hat{\mathbf{H}}\mathbf{D}_1\right)\right\} \quad (8)$$

*C. SVM Modulation Classification*

Feature-based modulation classification methods are of interest since they have lower complexity than likelihood-based methods. Here, we propose an SVM modulation classification method that combines multiple features.

The fourth order cumulants, $|C_{40}|$ and $C_{42}$ were previously proposed for modulation classification [10]. It is known that fourth order cumulants can be applied to distinguish between modulations, and are robust to noise effects. Given $M$ samples of a signal $s(k)$, cumulant $C_{40}$ is defined [10, 11]

$$C_{40} = \frac{\frac{1}{M}\sum_{k=0}^{M-1} s(k)^4 - 3\left(\frac{1}{M}\sum_{k=0}^{M-1} s(k)^2\right)^2}{\left(\frac{1}{M}\sum_{k=0}^{M-1}|s(k)|^2\right)^2} \quad (9)$$

whereas $C_{42}$ is defined

$$C_{42} = \frac{\frac{1}{M}\sum_{k=0}^{M-1} s(k)^4 - \left|\frac{1}{M}\sum_{k=0}^{M-1} s(k)^2\right|^2 - 2\left(\frac{1}{M}\sum_{k=0}^{M-1}|s(k)|^2\right)^2}{\left(\frac{1}{M}\sum_{k=0}^{M-1}|s(k)|^2\right)^2} \quad (10)$$

It is shown in [10] that $|C_{40}|$ and $C_{42}$ are invariant to carrier phase offset, which in our case corresponds to the phase ambiguity inherited from the ICA algorithm. Therefore, classification based on these features is not affected by the unknown phase offsets introduced by the ICA algorithm.

SVM is an important pattern recognition method, in which each pattern is represented by $D$ features [12-13]. The aim in SVM is to find the best separating hyperplane in $D$-dimensional space to discriminate between the patterns. SVM processing has two main steps: training and testing. In the training step, training data obtained from known sources is processed to find the optimum hyperplane separating data of different modulations. In the second stage of SVM, the test data is compared with the separating hyperplane and then classified accordingly. The training data in the proposed method corresponds to a number of $|C_{40}|$ and $C_{42}$ values for each candidate modulation and for each SNR value. Since thresholding between the modulations is SNR dependent, SVM modulation classification requires knowledge of the SNR at the receiver.

For modulation classification of MIMO signals, the signals $s(k)$ are the MIMO streams separated by ICA used one at a time in (9) and (10). It is interesting to compare the computational complexity of the two methods. The number of basic operations required for the SVM approach is $O(KN \times M_t)$, where $KN$ is the number of data samples processed in the ICA algorithm. In contrast, the likelihood-based approach requires $O(KN \times M_u^{M_t})$ basic operations, where $M_u$ is the maximum number of hypothetical states of the assumed constellations. For instance, if the problem is to classify between QPSK and 16QAM modulations, $M_u = 16$. Thus the complexity of the likelihood-based algorithm can be considerably higher than that of the SVM algorithm.

IV. UPPER BOUNDS ON THE PERFORMANCE OF MODULATION CLASSIFICATION

In the previous section, we have proposed an approach to classification of MIMO OFDM modulations relying on the flat nature of individual OFDM channels. In this section, we develop a CRB on the performance of channel estimation and upper bounds on performance of modulation classification for MIMO OFDM communication.

Modulation classification generally entails a step of blind channel estimation. As a result, the performance of modulation classification depends on the accuracy of the channel estimates [1], [5]. The well-known CRB provides a lower bound on the variance of any unbiased channel estimator. We evaluate the performance of modulation classification when the channel estimation error is provided by the CRB. Our approach can be viewed as an extension of [15] to the MIMO case. Specifically, we assume that the channel estimates are Gaussian random variables, with mean value equal to the true channel values, and variances equal to the CRB. An upper bound on the probability of correct classification is obtained by modulation classification (8), with channel estimates formed by Gaussian random variables with means equal to the true channel values and variances equal to the CRB.

The evaluation of the CRB for estimating the elements of the MIMO channel matrix is a key element in pursuing the upper bound on the performance of modulation classification. Next, the CRB for blind channel estimation is derived, followed by the upper bound on the performance of modulation classification.

*A. CRB of Channel Estimation*

*Data-Aided CRB*

Modulation classification is practiced with transmitted symbols that are, naturally, unknown to the receiver side. Similarly, the channel characteristics are not known to the receiver, and hence need to be estimated blindly. Errors in channel estimation lead to errors in modulation classification. A modulation classification bound obtained based on data-aided channel estimates is looser than the bound obtained based on blind channel estimation, yet may provide useful insight into other factors affecting classification performance. Here, we compute the CRB for the data-aided estimation of a flat MIMO channel.

Specifically, our goal is to derive the CRB for estimation of channel matrix $\mathbf{H}$ assuming the transmitted symbols $\mathbf{S}^{(i)}$ are known to the receiver. Let $\mathbf{h}$ denote the $2M_rM_t \times 1$ vector,

$$\mathbf{h} = \left[ \text{vec}\left(\mathbf{H}^{(R)}\right), \text{vec}\left(\mathbf{H}^{(I)}\right) \right]^T \quad (11)$$

where $\mathbf{H}^{(R)}$ and $\mathbf{H}^{(I)}$ are respectively, the real and imaginary parts of the channel matrix $\mathbf{H}$ and "vec" denotes the vectorizing operator, which means stacking the columns of a matrix. The CRB of $\mathbf{h}$ is expressed:

$$\text{CRB}(\mathbf{h}) = \mathbf{J}^{-1} \quad (12)$$

where $\mathbf{J}$ denotes the Fisher information matrix (FIM) of dimensions $2M_rM_t \times 2M_rM_t$. Each element of $\mathbf{J}$ depends on the expected value of the second-order partial derivatives with respect to the real or imaginary parts of the channel matrix elements. The matrix $\mathbf{J}$ is partitioned into 4 $M_rM_t \times M_rM_t$ blocks

$$\mathbf{J} = \begin{bmatrix} \mathbf{J}_{11} & \mathbf{J}_{12} \\ \mathbf{J}_{21} & \mathbf{J}_{22} \end{bmatrix}$$

We show the computation for the elements of the block $\mathbf{J}_{11}$. Similar derivations apply to other blocks. According to the vectorization in (11), the $\left[(n-1)M_r + m, (q-1)M_r + p\right]$ element of $\mathbf{J}_{11}$ is given by

$$\mathbf{J}_{11\left[(n-1)M_r+m,\, (q-1)M_r+p\right]} = -E\left[\frac{\partial^2 \ln L^{(i)}\left(\mathbf{Y} | \mathbf{S}^{(i)}, \mathbf{H}\right)}{\partial H_{mn}^{(R)} \partial H_{pq}^{(R)}}\right] \quad (13)$$

where $H_{ij}^{(R)}$ denotes the $(i, j)$ element of channel matrix $\mathbf{H}^{(R)}$. The observations comprising matrix $\mathbf{Y}$ are mutually independent (see (4)), therefore (13) is written

$$\mathbf{J}_{11\left[(n-1)M_r+m,\, (q-1)M_r+p\right]} = -\sum_{k=1}^{KN} E\left[\frac{\partial^2 \ln L\left(\mathbf{y}_k | \mathbf{s}_k, \mathbf{H}\right)}{\partial H_{mn}^{(R)} \partial H_{pq}^{(R)}}\right] \quad (14)$$

To compute the second derivative in (14), we start with the first derivative. By using (4), one can show that

$$\frac{\partial \ln L\left(\mathbf{y}_k | \mathbf{s}_k, \mathbf{H}\right)}{\partial H_{mn}^{(R)}} = -\frac{1}{\sigma^2} \frac{\partial \|\mathbf{y}_k - \mathbf{H}\mathbf{s}_k\|^2}{\partial H_{mn}^{(R)}} \quad (15)$$

Taking the partial derivative in (15) with respect to $H_{mn}^{(R)}$,

$$\frac{\partial \ln L\left(\mathbf{y}_k | \mathbf{s}_k, \mathbf{H}\right)}{\partial H_{mn}^{(R)}} = -\frac{1}{\sigma^2}$$

$$\times \frac{\partial\left\{\left[y_{k,m} - \sum_{j=1}^{M_t} H_{mj}s_{k,j}\right]\left[y_{k,m} - \sum_{j=1}^{M_t} H_{mj}s_{k,j}\right]^*\right\}}{\partial H_{mn}^{(R)}} \quad (16)$$

where $y_{k,m}$ denotes the $m$th element of the vector $\mathbf{y}_k$, and $s_{k,j}$ denotes the $j$th element of vector $\mathbf{s}_k$. After some manipulations,

$$\frac{\partial \ln L\left(\mathbf{y}_k | \mathbf{s}_k, \mathbf{H}\right)}{\partial H_{mn}^{(R)}} = -\frac{1}{\sigma^2} \times \quad (17)$$

$$\frac{\partial\left\{-2\left(s_{k,n}y_{k,m}^*\right)^{(R)} H_{mn}^{(R)} + 2\left(s_{k,n}\sum_{j\neq n}^{M_t} H_{m,j}^* s_{k,j}^*\right)^{(R)} H_{mn}^{(R)} + |s_{k,n}|^2 \left(H_{mn}^{(R)}\right)^2\right\}}{\partial H_{mn}^{(R)}}$$

Further, one can show that

$$\frac{\partial \ln L\left(\mathbf{y}_k | \mathbf{s}_k, \mathbf{H}\right)}{\partial H_{mn}^{(R)}} = -\frac{1}{\sigma^2} \times \quad (18)$$

$$\left\{-2\left(s_{k,n}y_{k,m}^*\right)^{(R)} + 2\left(s_{k,n}\sum_{j\neq n}^{M_t} H_{m,j}^* s_{k,j}^*\right)^{(R)} + 2|s_{k,n}|^2 \left(H_{mn}^{(R)}\right)\right\}$$

Taking the second derivative, we obtain

$$E\left[\frac{\partial^2 \ln L\left(\mathbf{y}_k | \mathbf{s}_k, \mathbf{H}\right)}{\partial H_{mn}^{(R)} \partial H_{pq}^{(R)}}\right] = \begin{cases} 0 & \text{if } p \neq m \\ -\frac{2}{\sigma^2}\left(s_{k,n}s_{k,q}^*\right)^{(R)} & \text{if } p = m \end{cases} \quad (19)$$

Then it is straightforward to obtain the FIM element $\mathbf{J}_{11\left[(n-1)M_r+m,\, (q-1)M_r+p\right]}$

$$\mathbf{J}_{11[(n-1)M_r+m,\ (q-1)M_r+p]} = \begin{cases} 0 & \text{if } p \neq m \\ \dfrac{2}{\sigma^2}\sum_{k=1}^{KN}\left(s_{k,n}s_{k,q}^*\right)^{(R)} & \text{if } p = m \end{cases} \quad (20)$$

The expressions of the elements of the different blocks constituting $\mathbf{J}$ are shown in Table I. The CRB of the channel can be obtained by using (12). From Table I, the CRB for the data-aided case depends only on the transmitted symbols and noise power, but not on the channel $\mathbf{H}$. It is shown in the Appendix A that the CRB for data-aided case is achievable.

The CRB (12) implies that the variance of the estimate of the $l$-th element of $\mathbf{h}$, $h_l$, is lower bounded as:

$$\text{var}(h)_l \geq [\mathbf{J}^{-1}]_{ll} \quad (21)$$

An upper bound on the performance of modulation classification exploiting (21), will be developed in sequel.

TABLE I
EXPRESSION FOR ELEMENTS OF $\mathbf{J}$

| Partial derivative variables | If $p = m$ and $q \neq n$.[#] | Otherwise |
|---|---|---|
| $H_{mn}^{(R)}, H_{pq}^{(R)}$ + | $\dfrac{2}{\sigma^2}\sum_{k=1}^{KN}\left(s_{k,n}s_{k,q}^*\right)^{(R)}$ | 0 |
| $H_{mn}^{(R)}, H_{pq}^{(I)}$ | $\dfrac{2}{\sigma^2}\sum_{k=1}^{KN}\left(s_{k,n}s_{k,q}^*\right)^{(I)}$ | 0 |
| $H_{mn}^{(I)}, H_{pq}^{(R)}$ | $\dfrac{2}{\sigma^2}\sum_{k=1}^{KN}\left(s_{k,q}s_{k,n}^*\right)^{(I)}$ | 0 |
| $H_{mn}^{(I)}, H_{pq}^{(I)}$ | $\dfrac{2}{\sigma^2}\sum_{k=1}^{KN}\left(s_{k,n}s_{k,q}^*\right)^{(R)}$ | 0 |

[#] The condition $q \neq n$ applies for the case of the second order partial derivative with respect to $(H_{mn}^{(R)}, H_{pq}^{(I)})$ or $(H_{mn}^{(I)}, H_{pq}^{(R)})$.

### B. Blind Channel Estimation CRB

Blind channel estimation is required for modulation classification of MIMO signals. In this part, we derive a CRB for blind channel estimation. For the blind case, the transmitted symbols $\mathbf{S}^{(i)}$ are unknown. When deriving the CRB, the major difference between data-aided estimation and blind channel estimation is that rather of computing the FIM from the likelihood function $L^{(i)}(\mathbf{Y}|\mathbf{S}^{(i)},\mathbf{H})$ conditioned on transmitted symbols $\mathbf{S}^{(i)}$ (see (13)), the FIM is evaluated from the likelihood $L^{(i)}(\mathbf{Y}|\mathbf{H})$ averaged over all symbols of the constellation $\Omega_i$,

$$L^{(i)}(\mathbf{Y}|\mathbf{H}) = E_{\Omega_i}\left[L^{(i)}(\mathbf{Y}|\mathbf{S}^{(i)},\mathbf{H})\right] \quad (22)$$

Substituting (4) in (22), we obtain an expression of the likelihood $L^{(i)}(\mathbf{y}_k|\mathbf{H})$, averaged over the constellation symbols,

$$L^{(i)}(\mathbf{Y}|\mathbf{H}) = \\ = \sum_{k=1}^{KN}\dfrac{1}{|\Omega_i|^{M_t}}\dfrac{1}{(\pi\sigma^2)^{M_r}}\sum_{\mathbf{s}\in\Omega_i^{M_t}}\exp\left[-\dfrac{1}{\sigma^2}\|\mathbf{y}_k - \mathbf{Hs}_k\|^2\right] \quad (23)$$

where $|\Omega_i|$ is the cardinality of the modulation $\Omega_i$.

Similar to the data-aided case, we show here the computation for the elements of the block $\mathbf{J}_{11}$. The $\left[(n-1)M_r+m,\ (q-1)M_r+p\right]$ element of $\mathbf{J}_{11}$ is given by

$$\mathbf{J}_{11[(n-1)M_r+m,\ (q-1)M_r+p]} = -E\left[\dfrac{\partial^2 \ln L^{(i)}(\mathbf{Y}|\mathbf{H})}{\partial H_{mn}^{(R)}\partial H_{pq}^{(R)}}\right] \quad (24)$$

Recall that the $M_r \times 1$ vectors $\{\mathbf{y}_k\}_{k=1}^{KN}$ (see (2)) and each of their components are independent and identically distributed, hence, one can drop the dependency on $k$, and write (24) as

$$\mathbf{J}_{11[(n-1)M_r+m,\ (q-1)M_r+p]} = -KNE\left[\dfrac{\partial^2 \ln L^{(i)}(\mathbf{y}|\mathbf{H})}{\partial H_{mn}^{(R)}\partial H_{pq}^{(R)}}\right] \quad (25)$$

where $L^{(i)}(\mathbf{y}|\mathbf{H}) =$

$$= \dfrac{1}{|\Omega_i|^{M_t}}\dfrac{1}{(\pi\sigma^2)^{M_r}}\sum_{\mathbf{s}\in\Omega_i^{M_t}}\exp\left[-\dfrac{1}{\sigma^2}\|\mathbf{y}-\mathbf{Hs}\|^2\right] \quad (26)$$

Detailed expressions for evaluating (25) are developed in Appendix B. From (23), the evaluation of (25) entails taking the log of a sum. For this reason, an analytical derivation for $\mathbf{J}_{11[(n-1)M_r+m,\ (q-1)M_r+p]}$ (and therefore the CRB) is intractable, but (25) can be evaluated numerically, by Monte Carlo methods. To be more specific, samples of $\mathbf{y}$ can be generated according to constellation $\Omega_i$, channel matrix $\mathbf{H}$ and the noise power. The expectation over $\mathbf{y}$ in (25) can be approximated by substituting samples of $\mathbf{y}$ into the expression of $\partial^2 \ln L^{(i)}(\mathbf{y}|\mathbf{H})/\partial H_{mn}^{(R)}\partial H_{pq}^{(R)}$, which is a function of $\mathbf{y}$ and $\mathbf{H}$, and then using the sample mean of $\partial^2 \ln L^{(i)}(\mathbf{y}|\mathbf{H})/\partial H_{mn}^{(R)}\partial H_{pq}^{(R)}$ to approximate the ensemble average. The CRB can then be obtained from $\text{CRB}(\mathbf{h}) = \mathbf{J}^{-1}$.

### C. Upper Bounds on the Performance of Modulation Classification

In the previous subsection, we developed the CRBs of channel estimation for both the data-aided and blind cases. We are ready now to derive the upper bounds on the modulation classification performance based on the errors provided by the CRBs. The accuracy of channel estimates determines the performance of modulation classification. Any unbiased channel estimator is lower bounded by CRB. Hence, we can view the probability of correct classification with channel estimates errors equal to the CRB, as an upper bound on the performance of modulation classification. Unfortunately, an analytical expression for the probability of correct classification could not be obtained. Numerical results based on Monte Carlo simulations are presented in the next section.

## V. NUMERICAL EXAMPLES

Simulations were carried out to demonstrate the proposed MIMO-OFDM modulation classification methods in standard channel models. The numerical examples obtained from simulations cover a spectrum of channel cases, and are compared to the upper bounds developed from the CRBs as discussed above.

*A. Simulation Setup*

The classification performance is first evaluated for an OFDM system, with and without Doppler. Then, the CRB of channel estimation is presented for time-invariant flat channels. Finally, the classification performance for both time-invariant flat MIMO channels and OFDM over time-variant channels is compared with the performance bounds derived from the CRB.

In the numerical examples, the goal was to discriminate between BPSK and QPSK modulations. A standard ITU Rayleigh-fading channel model was assumed for MIMO-OFDM transmission. The Rayleigh channel taps were normalized such that the expected value of the path gains' total power was 1. It implies that each element of matrix $\mathbf{H}(k,n)$ (and therefore $\mathbf{H}$) in (1) is a circular complex zero-mean Gaussian random variable with unit variance without Doppler effect. Hence, we evaluated the performance of channel estimation over flat Rayleigh fading channel, i.e., the element of $\mathbf{H}$ is also a circular complex zero-mean Gaussian random variable with unit variance. The SNR/sample was defined $10\log M_t/\sigma^2$. In all examples, the MIMO system was equipped with two transmit and four receive antennas, i.e., $M_t = 2$ and $M_r = 4$. The JADE algorithm was used to perform the blind channel estimation.

*B. Modulation Classification Performance for MIMO OFDM*

The OFDM system parameters are shown in Table II. Performance was evaluated over the ITU pedestrian B and vehicular A channel models, with the parameters given in Table III. The maximum speeds were assumed 3 km/hr for pedestrians, and 60 km/hr for vehicles. Time variations of the channels were modeled according to the Clarke and Gans model [14].

In our examples, we exploited the coherence bandwidth $B_c$ and time coherence $T_c$ [14] to increase the number of observations. These parameters are defined respectively, $B_c = 1/50\sigma_\tau$ and $T_c = 9/16\pi f_d$, where $\sigma_\tau$ is the RMS delay spread for the ITU channel models, and $f_d$ is the Doppler shift depending on carrier frequency and the vehicle speed. According to Table III, $\sigma_\tau$ and $B_c$ of Vehicular A channel model can be calculated to be 375.7nsec and 53.25kHz respectively. Therefore, a number of $B_c/\Delta f \approx 5$ subcarriers are within coherence bandwidth of ITU vehicular A channel. Similarly, by dividing the coherence time of the Vehicular A channel model by the OFDM frame duration, it can be calculated that 18 OFDM frames are within the coherence time for a maximum speed of 60 km/hr. In the numerical examples, we assume that the channel is fixed over 5 subcarriers and 10 OFDM frames. For each SNR value, the probability of correct classification was evaluated based on 500 Monte Carlo trials with different realizations of the pedestrian/vehicular channel model and of the noise.

TABLE II OFDM SYSTEM PARAMETERS

| Carrier frequency | 2 GHz |
|---|---|
| Total bandwidth | 6.4MHz |
| Number of subcarriers | 512 |
| Subcarrier spacing $\Delta f$ | 12.5kHz |
| Length of cyclic prefix | 32 |
| Sampling period | 156.25nsec |
| OFDM Frame duration (CP included) | 85usec |

TABLE III [21]
ITU TAPPED DELAY LINE CHANNEL PARAMETERS FOR PEDESTRIAN B AND VEHICULAR A TEST ENVIRONMENT

| Tap | Pedestrian B | | Vehicular A | | Doppler spectrum |
|---|---|---|---|---|---|
| | Relative delay (ns) | Average power (dB) | Relative delay (ns) | Average power (dB) | |
| 1 | 0 | 0 | 0 | 0.0 | Classic |
| 2 | 200 | –0.9 | 310 | -1.0 | Classic |
| 3 | 800 | –4.9 | 710 | -9.0 | Classic |
| 4 | 1 200 | –8.0 | 1090 | -10.0 | Classic |
| 5 | 2 300 | –7.8 | 1730 | -15.0 | Classic |
| 6 | 3 700 | –23.9 | 2510 | -20.0 | Classic |

In the training stage of the SVM, 100 symbols from each modulation BPSK and QPSK were generated synthetically. White Gaussian noise with a set variance corresponding to the SNR was added to the modulated signal. The features $|C_{40}|$ and $C_{42}$ of the noisy training data were calculated according to (9) and (10). SVM was implemented with the Matlab Bioinformatics Toolbox.

Fig. 3 shows the probability of correct classification for pedestrian B channels, and both stationary and moving terminals. In this figure, performance was evaluated using observations collected from 50 OFDM frames on one subcarrier, i.e., $K = 50$ and $N = 1$. Classification scores are shown for both likelihood-based and SVM-based methods. It is observed that both methods perform well for stationary and moving pedestrians. This is not surprising since at 3 km/hr, a pedestrian experiences a negligible Doppler spread. To achieve 85% correct classification rate, the likelihood-based approach outperformed the SVM approach by approximately 10 dB at the expense of extra computation complexity.

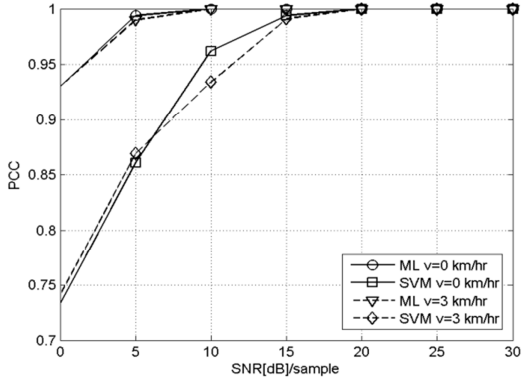

Fig. 3. PCC over ITU Pedestrian B channel for MIMO OFDM for classification of BPSK vs. QPSK.

In Fig. 4 are shown the successful classification rates for stationary and moving terminals for vehicles moving at 60 km/hr, the performance is significantly degraded. The degradation in classification performance for this case stems from the failed assumption that the channels remain static during the period of 50 OFDM frames. There exist large errors in estimating the time varying channel by using ICA.

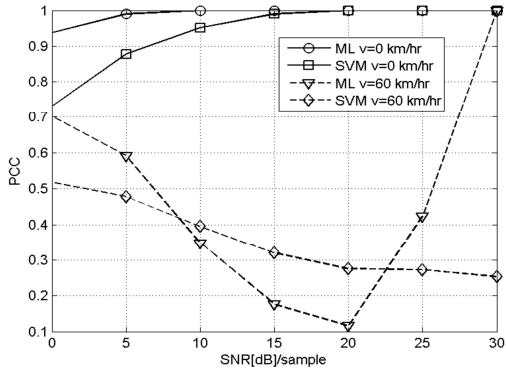

Fig. 4. Probability of correct classification over the ITU Vehicular A channel for classification of MIMO-OFDM BPSK vs. QPSK.

One natural solution to improve the performance over time varying channel is to limit the number of OFDM frames to match the coherence time of the channel. Regarding the OFDM system simulated, for a moving vehicle at the speed of 60 km/hr, the coherence time of ITU Vehicular channel A spans approximately 10 OFDM frames. But the received samples from 10 OFDM frames over one subcarrier are inadequate for ICA to perform successful blind channel estimation. To overcome the shortage in OFDM frames stemming from the relatively low channel time coherence, we exploit the similarity of OFDM channels within the coherence bandwidth. To this end, data is collected from groups of subcarriers within the same coherence bandwidth and groups of frames within the same coherent time. Hence, we grouped the data of 5 subcarriers and 10 frames in the simulations. The performance is shown in Fig. 5. This grouping method significantly improved the classification rate for moving vehicles at high speed. For example, at 60 km/hr, likelihood-based correct classification is close to guaranteed for SNR greater than 5 dB, while classification with SVM has a success rate higher than 80% for SNR greater than 5 dB.

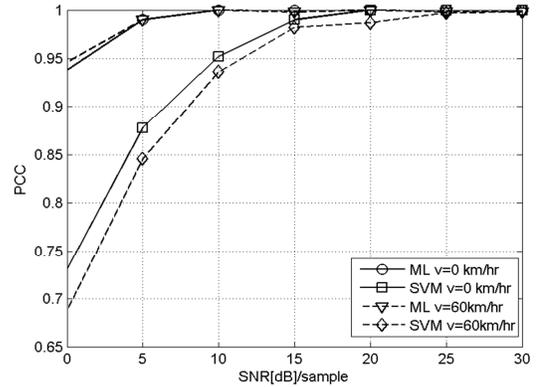

Fig. 5. Probability of correct classification over the ITU Vehicular A channel for classification of MIMO-OFDM BPSK vs. QPSK and grouping according to the coherence bandwidth and time.

*C. CRB of Channel Estimate*

The quantity $E\left[\partial^2 \ln L^{(i)}(\mathbf{y}|\mathbf{H})/\partial h_u \partial h_v\right]$ in (25) is required for the evaluation of the blind channel estimation CRB. Since an analytical expression is not available, (25) was evaluated from 1000 observations of the vector $\mathbf{y}$ (with different realizations of transmitted symbols and noise). The CRB shown in fig. 6 is an average of 500 CRBs computed for different instantiations of the channel $\mathbf{H}$. The modulation was assumed BPSK, and $KN$ in (25) was $KN=50$. Since a performance evaluation addressing the full channel matrix $\mathbf{H}$ is cumbersome, in the numerical examples, we focus on the channel estimation and CRB of $H_{11}^R$.

The ICA (JADE) algorithm is used for blind channel estimation. As discussed in Sections III.A and III.B, the channel estimate obtained from the JADE algorithm in conjunction with the phase correction (7) still leaves an unresolved phase ambiguity, which while not affecting modulation classification (the main topic of this paper), it will result in an error in the channel estimation. Since our focus is the effect of channel estimation on modulation classification, to obtain the results in Fig. 6, we corrected synthetically for the leftover phase ambiguity. Specifically, we obtain an estimate $\widehat{\mathbf{H}}$ of the channel matrix, correct the phase according to (7), and find a matrix $\mathbf{Q}$ product of an $M_t \times M_t$ permutation matrix and a diagonal leftover phase matrix such that $\mathbf{Q} = \min_{\mathbf{M}} \left\|\left(\widehat{\mathbf{H}}\mathbf{D}_1\mathbf{M}\right)^+ \mathbf{H} - \mathbf{I}\right\|$. The matrix $\mathbf{Q}$ is found by exhaustive search.

The mean square error (MSE) of $H_{11}^{(R)}$ versus SNR/sample is shown and compared to the CRB for the data-aided case and the blind channel estimation CRB in Fig. 6. It can be observed that three curves conflate at high SNR: Data-aided CRB, blind estimated CRB, and data-aided least squares channel estimation. Blind channel estimation by JADE exhibits a noise floor, which can be attributed to the imperfect separation of the MIMO signal streams.

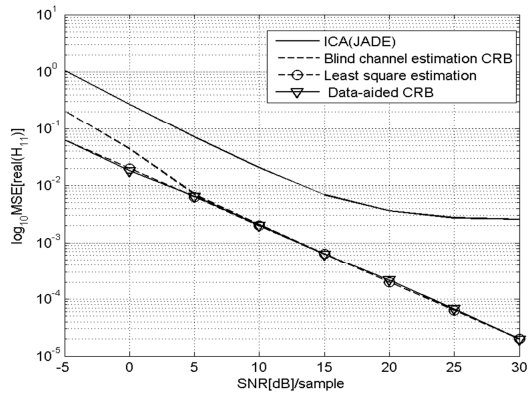

Fig. 6. Data-aided and blind channel estimation performance and CRBs of $H_{11}^{(R)}$.

## D. Upper Bounds on PCC

The approach here is to evaluate the probability of correct classification by Monte Carlo simulations in which the channel values were either estimated by ICA or simulated as the true values offset by random errors. The random errors were drawn from a Gaussian distribution with zero mean and variance equal to the CRB. Two CRB's were considered: data-aided (see Section IV.A) and blind (see Section IV.B). In practice, we expect channel estimation errors to be higher than the CRB, hence this approach leads to an upper bound on the performance of modulation classification. In all cases, classification was carried out by maximum likelihood according to (8). Note that (8) contains the phase offsets matrix $\mathbf{D}_1$. This matrix has to be evaluated when (8) is applied with channel estimates obtained from ICA. For the upper bound analysis, since, as previously discussed, we are using true channel values offset by random errors, $\mathbf{D}_1$ is taken as the identity matrix.

In Fig. 7, we denote PCC-DA the upper bound on the probability of correct classification derived with data-aided channel estimation, and PCC-B the upper bound derived with blind channel estimation, i.e., the channel estimate derived from the ICA. For each case, the figure shows the probability of correct classification computed as an average over probabilities of correct classification obtained for 500 different channel realizations. In addition to PCC-DA and PCC-B, Fig. 7 displays the PCC with known channel, the PCC for flat, stationary channels estimated by least squares, and the PCC for OFDM operating over ITU Vehicular A channels at 60 km/hr. The subcarriers of the OFDM signal are grouped according to the coherence bandwidth, and the frames are grouped according to the time coherence. It is observed in the figure, that as expected, the PCC with channel knowledge provides an upper bound of any modulation classification algorithm. The PCCs obtained from the CRBs for data-aided and blind channel estimation upper bounds on correct classification obtained from channel estimation with errors limited by the CRBs, are tighter (meaning closer to the actual performance) than the PCC with channel knowledge. Moreover, PCC-B is tighter than the PCC-DA. The PCC for the time-variant channel for OFDM is almost identical to the PCC obtained for the stationary, flat channel. This demonstrates the validity of our proposed approach of grouping subcarriers according to the coherence bandwidth and OFDM frames according to the time coherence. Fig. 7 also shows that to achieve 85% correct classification rate, the PCC-B bound outperformed the probability of correct classification for simulated MIMO data by only 2dB, which indicates that our proposed modulation classification approach for MIMO OFDM is quite close to the optimum.

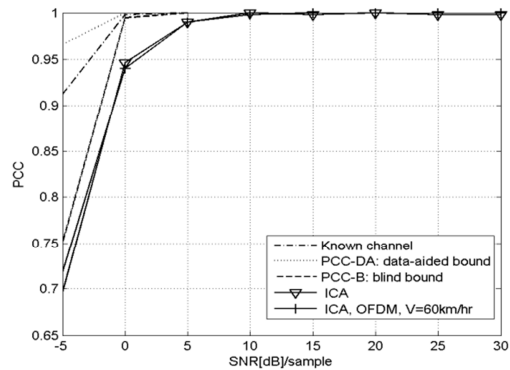

Fig. 7. Probability of correct classification for: known channel, data-aided CRB, blind CRB, flat stationary channel estimated by ICA, OFDM over Vehicular A channel estimated by ICA with coherence bandwidth and time coherence grouping.

## VI. CONCLUSIONS

This paper studies two classification modulation methods for MIMO-OFDM signals. First, the ICA JADE algorithm is applied to separate the data streams of the MIMO-OFDM signal. The modulation of the separated data streams is subsequently detected by ML or by SVM. Modulation classification of MIMO signals is more challenging than for SISO signals due to the residual multiuser interference resulting from imperfect signal separation. Our approach to modulation classification relies on the frequency non-selective nature of the channel experienced by individual OFDM data symbols, and exploits the invariance of the MIMO channel matrix across the coherence bandwidth and time coherence. In order to provide a benchmark against which we can compare the performance of our ML approach for flat MIMO channels, we derived two upper bounds (PCC-DA and PCC-B) on PCC based on the errors provided by the CRBs of the flat MIMO channel estimate. The proposed modulation classification methods for MIMO OFDM were demonstrated over ITU channels. Over the slowly time-varying ITU pedestrian channel, the proposed ML method achieved 95% classification rate for SNR/sample higher than 5 dB. The SVM approach attained 85% classification rate with lower complexity than the maximum likelihood approach, for SNR/sample higher than 5 dB. Over the fast fading channel, high probabilities of correct classification were maintained by grouping signals according to the coherent bandwidth and time coherence of the channel. In this case, performance was close to the static channel. It was shown by numerical simulations that proposed performance bounds on modulation classification derived based on CRBs are tighter than the bound with known channel knowledge and that the performance of ML classification is quite close to the optimum.

## VII. APPENDIX A

We prove here that the CRB of channel estimation for the data-aided case is achievable. Towards this objective, let us first recall the following theorem from [17]:

*Theorem A.1* (Theorem 4.1 in [17]): If the data observed can be modeled as

$$\mathbf{x} = \mathbf{B}\boldsymbol{\theta} + \mathbf{w}, \quad (A.1)$$

where $\mathbf{x}$ is a real-valued $M \times 1$ vector of observations, $\mathbf{B}$ is a known $M \times n$ real matrix (with $M > n$) and rank $n$, $\boldsymbol{\theta}$ is a real-valued $n \times 1$ vector of parameters to be estimated, and $\mathbf{w}$ is an $M \times 1$ noise vector with PDF $\mathcal{N}(\mathbf{0}, \sigma_\mathbf{w}^2 \mathbf{I})$, then an efficient estimator exists.

We need to prove that the theorem applies to our complex-valued model (2). To this end, we need to rewrite (2) in terms of real values only. Let $y_{k,m}$ be the $k$-th received sample at the $m$-th antenna, $y_{k,m}^{(R)}$ its real part, and $y_{k,m}^{(I)}$ its imaginary part. Let $\mathbf{h}_m$ be the $m$-th row of $\mathbf{H}$. Then, we have from (2),

$$y_{k,m} = \mathbf{h}_m \mathbf{s}_k + z_{k,m} \quad (A.2)$$

where $z_{k,m}$ is the noise component. We can express (A.2) as

$$\begin{bmatrix} y_{k,m}^{(R)} \\ y_{k,m}^{(I)} \end{bmatrix} = \begin{bmatrix} \left[\mathbf{s}_k^T\right]^{(R)} & -\left[\mathbf{s}_k^T\right]^{(I)} \\ \left[\mathbf{s}_k^T\right]^{(I)} & \left[\mathbf{s}_k^T\right]^{(R)} \end{bmatrix} \begin{bmatrix} \left[\mathbf{h}_m^T\right]^{(R)} \\ \left[\mathbf{h}_m^T\right]^{(I)} \end{bmatrix} + \begin{bmatrix} z_{k,m}^{(R)} \\ z_{k,m}^{(I)} \end{bmatrix} \quad (A.3)$$

Then extending (A.3) for $k = 1 \ldots KN$, we get

$$\begin{bmatrix} y_{1,m}^{(R)} \\ y_{1,m}^{(I)} \\ \vdots \\ y_{KN,m}^{(R)} \\ y_{KN,m}^{(I)} \end{bmatrix} = \begin{bmatrix} \left[\mathbf{s}_1^T\right]^{(R)} & -\left[\mathbf{s}_1^T\right]^{(I)} \\ \left[\mathbf{s}_1^T\right]^{(I)} & \left[\mathbf{s}_1^T\right]^{(R)} \\ \vdots & \\ \left[\mathbf{s}_{KN}^T\right]^{(R)} & -\left[\mathbf{s}_{KN}^T\right]^{(I)} \\ \left[\mathbf{s}_{KN}^T\right]^{(I)} & \left[\mathbf{s}_{KN}^T\right]^{(R)} \end{bmatrix} \begin{bmatrix} \left[\mathbf{h}_m^T\right]^{(R)} \\ \left[\mathbf{h}_m^T\right]^{(I)} \end{bmatrix} + \begin{bmatrix} z_{1,m}^{(R)} \\ z_{1,m}^{(I)} \\ \vdots \\ z_{KN,m}^{(R)} \\ z_{KN,m}^{(I)} \end{bmatrix} \quad (A4)$$

Signal model (A.5) is now in the form of (A.1) with $\boldsymbol{\theta} = \begin{bmatrix} \mathbf{h}_m^{(R)} & \mathbf{h}_m^{(I)} \end{bmatrix}^T$.

## VIII. APPENDIX B
EXPRESSIONS OF $\partial^2 \ln L^{(i)}(\mathbf{y} | \mathbf{H}) / \partial H_{mn}^{(R)} \partial H_{pq}^{(R)}$

In this appendix, we develop expressions for evaluating (25).

We derive the expression for $\partial^2 \ln L^{(i)}(\mathbf{y} | \mathbf{H}) / \partial H_{mn}^{(R)} \partial H_{qp}^{(R)}$. The expressions for other second order derivatives, $\partial^2 \ln L^{(i)}(\mathbf{y} | \mathbf{H}) / \partial H_{mn}^{(R)} \partial H_{pq}^{(I)}$, $\partial^2 \ln L^{(i)}(\mathbf{y} | \mathbf{H}) / \partial H_{mn}^{(I)} \partial H_{pq}^{(R)}$ and $\partial^2 \ln L^{(i)}(\mathbf{y} | \mathbf{H}) / \partial H_{mn}^{(I)} \partial H_{pq}^{(I)}$ can be obtained similarly. By using (26) and applying the chain rule, one can show that

$$\frac{\partial \ln L^{(i)}(\mathbf{y} | \mathbf{H})}{\partial H_{mn}^{(R)}} = \left\{ \sum_{\mathbf{s} \in \Omega_i^{M_t}} \exp\left[-\sigma^{-2} \|\mathbf{y} - \mathbf{H}\mathbf{s}\|^2\right] \right\}^{-1} \times$$

$$\times \sum_{\mathbf{s} \in \Omega_i^{M_t}} \frac{\partial \exp\left[-\sigma^{-2} \|\mathbf{y} - \mathbf{H}\mathbf{s}\|^2\right]}{\partial H_{mn}^{(R)}}$$

Then, according to the product rule and the chain rule of differentiation

$$\frac{\partial^2 \ln L^{(i)}(\mathbf{y} | \mathbf{H})}{\partial H_{mn}^{(R)} \partial H_{pq}^{(R)}} = -\left\{ \sum_{\mathbf{s} \in \Omega_i^{M_t}} \exp\left[-\sigma^{-2} \|\mathbf{y} - \mathbf{H}\mathbf{s}\|^2\right] \right\}^{-2} \times$$

$$\times \sum_{\mathbf{s} \in \Omega_i^{M_t}} \frac{\partial \exp\left[-\sigma^{-2} \|\mathbf{y} - \mathbf{H}\mathbf{s}\|^2\right]}{\partial H_{pq}^{(R)}} \times \sum_{\mathbf{s} \in \Omega_i^{M_t}} \frac{\partial \exp\left[-\sigma^{-2} \|\mathbf{y} - \mathbf{H}\mathbf{s}\|^2\right]}{\partial H_{mn}^{(R)}} +$$

$$+ \left\{ \sum_{\mathbf{s} \in \Omega_i^{M_t}} \exp\left[-\sigma^{-2} \|\mathbf{y} - \mathbf{H}\mathbf{s}\|^2\right] \right\}^{-1} \times \sum_{\mathbf{s} \in \Omega_i^{M_t}} \frac{\partial^2 \exp\left[-\sigma^{-2} \|\mathbf{y} - \mathbf{H}\mathbf{s}\|^2\right]}{\partial H_{mn}^{(R)} \partial H_{pq}^{(R)}}$$

(B.1)

It can be seen from (B.1) that, we need to compute

$$\partial \exp\left[-\sigma^{-2} \|\mathbf{y} - \mathbf{H}\mathbf{s}\|^2\right] / \partial H_{mn}^{(R)}$$

and $\quad \partial^2 \exp\left[-\sigma^{-2} \|\mathbf{y} - \mathbf{H}\mathbf{s}\|^2\right] / \partial H_{mn}^{(R)} \partial H_{pq}^{(R)}$.

Applying the chain rule again,

$$\frac{\partial \exp\left[-\sigma^{-2} \|\mathbf{y} - \mathbf{H}\mathbf{s}\|^2\right]}{\partial H_{mn}^{(R)}}$$

$$= \exp\left[-\sigma^{-2} \|\mathbf{y} - \mathbf{H}\mathbf{s}\|^2\right] \times \frac{\partial \left[-\sigma^{-2} \|\mathbf{y} - \mathbf{H}\mathbf{s}\|^2\right]}{\partial H_{mn}^{(R)}} \quad (B.2)$$

Applying the product rule and chain rule of differentiation, we get

$$\frac{\partial^2 \exp\left[-\sigma^{-2} \|\mathbf{y} - \mathbf{H}\mathbf{s}\|^2\right]}{\partial H_{mn}^{(R)} \partial H_{pq}^{(R)}}$$

$$= \frac{\partial \exp\left[-\sigma^{-2} \|\mathbf{y} - \mathbf{H}\mathbf{s}\|^2\right]}{\partial H_{pq}^{(R)}} \frac{\partial \left[-\sigma^{-2} \|\mathbf{y} - \mathbf{H}\mathbf{s}\|^2\right]}{\partial H_{mn}^{(R)}}$$

$$+ \exp\left[-\sigma^{-2} \|\mathbf{y} - \mathbf{H}\mathbf{s}\|^2\right] \frac{\partial^2 \left[-\sigma^{-2} \|\mathbf{y} - \mathbf{H}\mathbf{s}\|^2\right]}{\partial H_{mn}^{(R)} \partial H_{pq}^{(R)}} \quad (B.3)$$

Since according to (15), we have

$$\frac{\partial \left[-\sigma^{-2} \|\mathbf{y} - \mathbf{H}\mathbf{s}\|^2\right]}{\partial H_{mn}^{(R)}} = \frac{\partial \ln L(\mathbf{y} | \mathbf{s}, \mathbf{H})}{\partial H_{mn}^{(R)}}$$

and $\quad \dfrac{\partial^2 \left[-\sigma^{-2} \|\mathbf{y} - \mathbf{H}\mathbf{s}\|^2\right]}{\partial H_{mn}^{(R)} \partial H_{pq}^{(R)}} = \dfrac{\partial^2 \ln L(\mathbf{y} | \mathbf{s}, \mathbf{H})}{\partial H_{mn}^{(R)} \partial H_{pq}^{(R)}}$

The terms $\partial \left[-\sigma^{-2} \|\mathbf{y} - \mathbf{H}\mathbf{s}\|^2\right] / \partial H_{mn}^{(R)}$ and

$$\partial^2 \left[-\sigma^{-2} \|\mathbf{y} - \mathbf{H}\mathbf{s}\|^2\right] / \partial H_{mn}^{(R)} \partial H_{pq}^{(R)}$$

in (B.2) and (B.3) can be obtained by using (18) and (19).

For other cases of the second order derivative, including $\partial^2 \ln L^{(i)}(\mathbf{y} | \mathbf{H}) / \partial H_{mn}^{(R)} \partial H_{pq}^{(I)}$, $\partial^2 \ln L^{(i)}(\mathbf{y} | \mathbf{H}) / \partial H_{mn}^{(I)} \partial H_{pq}^{(R)}$ and $\partial^2 \ln L^{(i)}(\mathbf{y} | \mathbf{H}) / \partial H_{mn}^{(I)} \partial H_{pq}^{(I)}$, one may compute them by applying the product rule and the chain rule of differentiation

similarly, and substituting the corresponding equations (B.4)-(B.9).

$$\frac{\partial\left[-\sigma^{-2}\|\mathbf{y}-\mathbf{Hs}\|^2\right]}{\partial H_{mn}^{(R)}} = -\frac{1}{\sigma^2} \times \\ \times\left\{-2\left(s_n y_m^*\right)^{(R)} + 2\left(s_n \sum_{j\neq n}^{M_t} H_{m,j}^* s_j^*\right)^{(R)} + 2|s_n|^2 \left(H_{mn}^{(R)}\right)\right\} \quad (B.4)$$

where we use $y_m$ to denote the *m*-th element of **y** and $s_n$ to denote the *n*-th element of **s**.

$$\frac{\partial\left[-\sigma^{-2}\|\mathbf{y}-\mathbf{Hs}\|^2\right]}{\partial H_{mn}^{(I)}} = -\frac{1}{\sigma^2} \times \\ \times\left\{2\left(s_n y_m^*\right)^{(I)} - 2\left(s_n \sum_{j\neq n}^{M_t} H_{m,j}^* s_j^*\right)^{(I)} + 2|s_n|^2 \left(H_{mn}^{(I)}\right)\right\} \quad (B.5)$$

$$\frac{\partial^2\left[-\sigma^{-2}\|\mathbf{y}-\mathbf{Hs}\|^2\right]}{\partial H_{mn}^{(R)} \partial H_{pq}^{(R)}} = -\frac{2}{\sigma^2}\left(s_n s_q^*\right)^{(R)} \quad (B.6)$$

$$\frac{\partial^2\left[-\sigma^{-2}\|\mathbf{y}-\mathbf{Hs}\|^2\right]}{\partial H_{mn}^{(R)} \partial H_{pq}^{(I)}} = -\frac{2}{\sigma^2}\left(s_n s_q^*\right)^{(I)} \quad (B.7)$$

$$\frac{\partial^2\left[-\sigma^{-2}\|\mathbf{y}-\mathbf{Hs}\|^2\right]}{\partial H_{mn}^{(I)} \partial H_{pq}^{(I)}} = -\frac{2}{\sigma^2}\left(s_q s_n^*\right)^{(I)} \quad (B.8)$$

$$\frac{\partial^2\left[-\sigma^{-2}\|\mathbf{y}-\mathbf{Hs}\|^2\right]}{\partial H_{mn}^{(I)} \partial H_{pq}^{(I)}} = -\frac{2}{\sigma^2}\left(s_n s_q^*\right)^{(R)} \quad (B.9)$$